\documentclass[aps,prd,reprint,preprintnumbers,superscriptaddress,nofootinbib,floatfix,nolongbibliography]{revtex4-2}
\usepackage{amsmath,amssymb,bm}
\usepackage{array,booktabs}
\usepackage{graphicx}
\usepackage{microtype}
\usepackage{xcolor}
\usepackage[colorlinks=true,allcolors=blue!55!black]{hyperref}

\graphicspath{{Figures/}}
\hypersetup{
  pdftitle={Interpreting the LZ 248 keV Event using Dark QCD},
  pdfauthor={Francesco Sannino and Jessica Turner}
}
\newcommand{\DMl}{\chi_1}
\newcommand{\DMh}{\chi_2}
\newcommand{\ii}{\mathrm{i}}

\begin{document}
\preprint{IPPP/26/65}
\title{Interpreting the LZ 248 keV Event using Dark QCD}

\author{Francesco Sannino}
\email{sannino@qtc.sdu.dk}
\affiliation{Quantum Theory Centre ($\hbar$QTC),
University of Southern Denmark, Campusvej 55,
DK-5230 Odense M, Denmark}
\affiliation{Department of Physics E. Pancini, Universit\`a di Napoli
Federico II, via Cintia, 80126 Napoli, Italy}
\affiliation{INFN Sezione di Napoli, via Cintia, 80126 Napoli, Italy}
\affiliation{Institute for Particle Physics Phenomenology, Durham University,
Durham DH1 3LE, United Kingdom}

\author{Jessica Turner}
\email{jessica.turner@durham.ac.uk}
\affiliation{Institute for Particle Physics Phenomenology, Durham University,
Durham DH1 3LE, United Kingdom}

\date{\today}

\begin{abstract}
The LUX–ZEPLIN (LZ) Collaboration has reported a nuclear-recoil candidate with reconstructed energy $248\pm23_{\rm stat}\pm23_{\rm sys}\,\mathrm{keV}$. We investigate an interpretation in terms of composite inelastic dark matter, in which a dimension-ten operator splits the lightest baryon of a confining $SU(3)_{\rm D}$ theory into two Majorana states. Endothermic scattering is mediated by a leptophobic vector coupled to baryon number. For a dark-baryon mass of $1\,\mathrm{TeV}$ and a splitting of $300\,\mathrm{keV}$, a reference nucleon cross section of $6.5\times10^{-43}\,\mathrm{cm^2}$ lies within LZ’s published two-sided 90\% confidence interval. This benchmark fixes $M_V/g_B\simeq9.9\,\mathrm{TeV}$, where $M_V$ and $g_B$ are the mediator mass and baryon-number gauge coupling, and is compatible with solar-neutrino and dijet constraints. 
\end{abstract}
\maketitle

\section{Introduction}
\label{sec:introduction}
The LUX--ZEPLIN (LZ) Collaboration has extended its nuclear-recoil (NR) analysis to approximately
$270\,\mathrm{keV}$ and reported one event in the upper
part of that range \cite{LZ:2026axp}.  Interpreted with the elastic-NR response,
its reconstructed recoil energy is
$E_{\rm obs}=248\pm23_{\rm stat}\pm23_{\rm sys}\,\mathrm{keV}$.
The event does not establish a signal: the largest local background-only
tension in the LZ model scan is $3.4\sigma$ and reduces to $2.6\sigma$ after the
look-elsewhere correction \cite{LZ:2026axp}.  Its large recoil energy motivates dark matter models with
suppressed low-energy recoils.
Endothermic scattering, $\DMl N\to\DMh N$ with
$m_2-m_1\equiv\delta>0$,
naturally suppresses low-energy recoils because the incident particle must
supply the excitation energy.  This mechanism and its preference for heavy
targets are well known \cite{Tucker-Smith:2001myb,Tucker-Smith:2004mxa}, and searches at high recoil energies were proposed before the present LZ result
\cite{Bramante:2016rdh,Barello:2014uda}.  Recent studies of the LZ event
have considered endothermic WIMPs, Higgsinos and strong seasonal modulation
\cite{Su:2026rwz,DiMauro:2026ldr,Fan:2026kxx,Rodd:2026tyn,McCabe:2026crm},
as well as dark photon and absorption mechanisms
\cite{Yamashita:2026ump,Lou:2026idn}.

Technibaryon dark matter has been studied in walking and ultra-minimal
technicolor theories \cite{Gudnason:2006ug,Gudnason:2006yj,Ryttov:2008xe},
with direct-detection and collider phenomenology developed in
Ref.~\cite{Foadi:2008qv}. Quasi-stable technibaryons with a baryon-like
asymmetry were also proposed to account for cosmic-ray electron and
positron excesses \cite{Nardi:2008ix}. Here the confining sector does not
break electroweak symmetry, and the dark-matter ground state is stable.
Ref.~\cite{Asadi:2026iot} also considers composite dark baryons,
but scattering proceeds through a magnetic
transition \cite{Asadi:2024bbq}. Ref.~\cite{Du:2026lpa} instead shares our
pseudo-Dirac structure and leptophobic baryon-number vector portal, but its
dark matter is elementary. In our construction, dark quarks that are Standard Model (SM) singlets confine
into a Dirac baryon. Two further complex scalar fields break the gauged
$U(1)_{\rm B}$ spontaneously, and one of them enters a higher-dimensional
operator that explicitly breaks the accidental dark baryon number and
splits the dark Dirac baryon into two Majorana states. For a TeV dark confinement scale and order-one Wilson and hadronic
coefficients, a $300\,\mathrm{keV}$ splitting corresponds to a suppression
scale of order $10\,\mathrm{TeV}$.
We compare a model benchmark with the published LZ interval and discuss
the corresponding collider searches. We also discuss solar neutrino
constraints \cite{Pospelov:2026sol,Nguyen:2026lui}, whose strength depends
on the dark hadron spectrum and its decays.
We treat the local dark baryon density as an input rather than deriving
the cosmological production of the dark matter; the fate of the excited
state is discussed in Sec.~\ref{sec:constraints}. Our comparison
uses the LZ exposure of $2.84$ tonne-years, the published energy-dependent
nuclear-recoil acceptance and the reconstructed recoil energy with its quoted
uncertainties \cite{LZ:2026axp}. A full detector-response model and an independent background
likelihood are beyond the scope of this paper.

\section{Composite pseudo-Dirac baryon as dark matter}
\label{sec:model}
\subsection{Field content and anomaly cancellation}\label{sec:gaugestr}
We extend the SM gauge group to
\begin{equation}
 G=G_{\rm SM}\times SU(3)_{\rm D}\times U(1)_{\rm B}\,,
 \label{eq:gauge-group}
\end{equation}
where the dark group $SU(3)_{\rm D}$ has coupling $g_{\rm D}$ and
confinement scale $\Lambda_{\rm D}$, and the gauged $U(1)_{\rm B}$ has
gauge boson $V_\mu$ and coupling $g_B$. We denote the gauged baryon charge
of a field $X$ by $Q_B(X)$: SM quarks have $Q_B=1/3$, while SM leptons and
the Higgs doublet $H$ have $Q_B=0$.

The two dark quarks $U$ and $D$ are
vector-like Dirac fermions, singlets under the SM, with
$(SU(3)_{\rm D},Q_B)=(\mathbf3,5/2)$. The gauged $U(1)_{\rm B}$ is broken by
two new complex SM singlet scalars, $S$ and $\Phi$, with baryon charges
$-15$ and $-3$, respectively. The vacuum expectation value of $\Phi$ breaks $U(1)_{\rm B}$ and
gives masses to the spectator fermions introduced below. The field $S$ enables
a dimension-ten operator which splits the dark baryon mass, and its
vacuum expectation value dominates the $U(1)_{B}$ vector mass at our benchmark. Assigning the accidental dark number
$N_{\rm D}=1/3$ to each dark quark, and zero to the other elementary
fields, gives $N_{\rm D}=1$ for a three-dark-quark baryon. This global
number is conserved by the renormalisable interactions but will be broken
by the same $S$-induced higher-dimensional operator.

The dark quarks do not contribute to gauge anomalies. To cancel those of gauged
SM baryon number, we add the colour- and dark-colour-singlet spectators
\cite{FileviezPerez:2018jmr,FileviezPerez:2019jju}
\begin{equation}
\begin{aligned}
 \Psi_L&\sim(\mathbf2,-\tfrac12,B_1),&
 \Psi_R&\sim(\mathbf2,-\tfrac12,B_2)\,,\\
 \eta_R&\sim(\mathbf1,-1,B_1),&
 \eta_L&\sim(\mathbf1,-1,B_2)\,,\\
 N_R&\sim(\mathbf1,0,B_1),&
 N_L&\sim(\mathbf1,0,B_2)\,,
\end{aligned}
\label{eq:spectators}
\end{equation}
where the entries denote the $SU(2)_L$ representation, the hypercharge $Y$
and the baryon charge $Q_B$, and $\Psi_{L,R}=(\Psi^0,\Psi^-)_{L,R}^T$. These fermions are vector-like
under the SM but chiral under $U(1)_{\rm B}$. Anomaly cancellation requires
\begin{equation}
\begin{aligned}
 \mathcal A_{SU(2)^2B}&=\frac32+\frac{B_1-B_2}{2}=0\,,\\
 \mathcal A_{Y^2B}&=-\frac32-\frac{B_1-B_2}{2}=0\,.
\end{aligned}
\label{eq:anomaly-check}
\end{equation}
We select $(B_1,B_2)=(-1,2)$. The neutral singlets $N_{L,R}$ do not
enter these two triangles, but are required to cancel the cubic
$U(1)_{\rm B}$ and mixed gravitational anomalies.

\subsection{Symmetry breaking and spectator masses}\label{sec:symbreak}
The dark-sector Lagrangian is
\begin{align}
 \mathcal L_{\rm D}={}&-\frac14G^A_{\mu\nu}G^{A\mu\nu}
 -\frac14V_{\mu\nu}V^{\mu\nu}\nonumber\\
 &+\sum_{Q=U,D}\bar Q(\ii\gamma^\mu D_\mu-m_Q)Q
 +|D_\mu S|^2+|D_\mu\Phi|^2\nonumber\\
 &-V(S,\Phi,H)+\mathcal L_{\Delta N_{\rm D}=2}\,,
 \label{eq:lagrangian}
\end{align}
where $G^A_{\mu\nu}$, $A=1,\ldots,8$, and $V_{\mu\nu}$ are the
$SU(3)_{\rm D}$ and $U(1)_{\rm B}$ field strengths, $m_U,m_D$ are the
dark-quark masses, $V(S,\Phi,H)$ is the scalar potential, specified in
Eq.~\eqref{eq:phase-locking} and $\mathcal L_{\Delta N_{\rm D}=2}$ is the
dark-number-violating operator of Eq.~\eqref{eq:splitting-operator}. For
the SM singlet fields the covariant derivative is
\begin{equation}
 D_\mu X=(\partial_\mu-\ii g_{\rm D}T^AG^A_\mu
                 -\ii g_BQ_B(X)V_\mu)X\,,
 \label{eq:covariant-derivative}
\end{equation}
with $T^A$ the $SU(3)_{\rm D}$ generators, normalised as
$\operatorname{tr}(T^AT^B)=\delta^{AB}/2$ in the fundamental. The vector couples to SM quarks with $g_q=g_B/3$
and has no tree-level SM-lepton coupling when kinetic mixing is neglected.

The scalars acquire vacuum expectation values
$\langle S\rangle=v_S/\sqrt2$ and $\langle\Phi\rangle=v_\Phi/\sqrt2$, with
real positive $v_S,v_\Phi$, breaking $U(1)_{\rm B}$ and giving $V_\mu$ the
mass squared
\begin{equation}
 M_V^2=g_B^2(225v_S^2+9v_\Phi^2)\,.
 \label{eq:vector-mass}
\end{equation}
The spectator Yukawa interactions are
\begin{align}
 \mathcal L_{\rm spec}\supset{}&
 -y_\Psi\bar\Psi_L\Phi\Psi_R
 -y_\eta\bar\eta_L\Phi^\dagger\eta_R\nonumber\\
 &-y_N\bar N_L\Phi^\dagger N_R+\mathrm{h.c.}
 \label{eq:spectator-masses}
\end{align}
Gauge invariance of the first term fixes
$Q_B(\Phi)=B_1-B_2=-3$.
Before mixing induced by electroweak symmetry breaking, the spectator masses are
$m_i=|y_i|v_\Phi/\sqrt2$, $i=\Psi,\eta,N$.
The baryon charges of $N_R$ forbid a neutrino-portal Yukawa $\bar L\widetilde HN_R$, where $L$ is an SM lepton doublet and $\widetilde H=\ii\sigma_2H^*$.
Allowed Higgs couplings mix $\Psi^0$ with $N$ and $\Psi^-$ with $\eta$. We take the lightest
spectator to be neutral, so that the charged states decay through
Higgs-induced mixing, and we introduce a spectator-number-violating
operator that prevents the
neutral spectator from remaining as a stable relic.

The scalar potential is
\begin{equation}
 V=V_{\rm ren}
 -\left(\frac{\kappa_\Phi}{\Lambda_\Phi^2}S^\dagger\Phi^5
       +\mathrm{h.c.}\right)\,,
 \label{eq:phase-locking}
\end{equation}
where $V_{\rm ren}$ contains all renormalisable scalar interactions of $\Phi$, $S$ and $H$, $\kappa_\Phi$ is dimensionless and $\Lambda_\Phi$ is a heavy
scale. This dimension-six interaction is the lowest-dimensional scalar
term that induces a mass for the pseudoscalar $a$ not absorbed in the longitudinal component of $V$. Taking $\kappa_\Phi$ real and positive,
\begin{equation}
 m_a^2=\frac{\kappa_\Phi v_Sv_\Phi^5}{4\Lambda_\Phi^2}
 \left(\frac1{v_S^2}+\frac{25}{v_\Phi^2}\right)\,.
 \label{eq:pseudoscalar-mass}
\end{equation}
For $\kappa_\Phi$ of order one and $\Lambda_\Phi$ of order $10\,\mathrm{TeV}$, $m_a$ is of order $100\,\mathrm{GeV}$; a strongly suppressed $\kappa_\Phi$ instead leaves $a$ light, which is the case relevant for the depletion of the excited baryon discussed in Sec.~\ref{sec:constraints}.

Rescaling all baryon charges by six so that they are integers, $\widehat Q_B=6Q_B$, the scalar charges of $S$ and $\Phi$ are $-90$ and $-18$, respectively. For the gauge group in
Eq.~\eqref{eq:gauge-group} their greatest common divisor leaves a
$\mathbb{Z}_{18}$ gauge subgroup, whose order-two element acts as
\begin{equation}
 P(X)=(-1)^{\widehat Q_B(X)}\,.
 \label{eq:dark-parity}
\end{equation}
Both condensates, the SM fields and the spectators are even, whereas
each dark quark has $\widehat Q_B=15$ and is odd. The lightest
three-dark-quark baryon is consequently stable. The $\mathbb{Z}_{18}$ residual symmetry also
forbids SM-only operators with $\Delta B_{\rm SM}=1,2$, including the
usual proton-decay and neutron--antineutron operators, since the scalar vacuum
can only supply baryon charge in multiples of three. We note that if the breaking $U(1)_{\rm B}\to\mathbb{Z}_{18}$ occurs after inflation, it can produce a network of local cosmic strings.

\subsection{Confinement and the Majorana splitting}\label{sec:majorana}
For $m_U,m_D\lesssim\Lambda_{\rm D}$, we assume that the lightest dark
baryon is a spin-$1/2$ state $B\sim UUD$, with
$m_B=\mathcal O(\Lambda_{\rm D})$ and $Q_B(B)=15/2$.
A small positive $m_D-m_U$
favours $UUD$ over $UDD$.

Dark mesons $\bar QQ$ have $Q_B=0$ and are even under the stabilising
parity. For $m_U,m_D\ll\Lambda_{\rm D}$, the lightest are three
pseudo-Goldstone pions, which are electrically neutral.
One way to make all three pions unstable, and decay to SM quarks, is to introduce a dimension-six interaction of the type:
\begin{equation}
 \mathcal L_\pi=\frac{1}{\Lambda_\pi^2}
 \sum_{q=d,s,b}\left(\bar Q\gamma_\mu\gamma_5 C_qQ\right)
                    \left(\bar q_R\gamma^\mu q_R\right)\,,
 \label{eq:pion-decay}
\end{equation}
where $Q=(U,D)^T$ is the dark-flavour doublet with dark-colour indices
contracted, $q_R$ are the right-handed SM down-type quark fields, the $C_q$
are Hermitian $2\times2$ matrices in dark-flavour space and $\Lambda_\pi$ is
the scale of the interaction. 
The interaction is diagonal in SM flavour and introduces no tree-level
SM flavour-changing neutral currents, and it respects the
gauge symmetry, the dark parity and dark number.

An interpolating operator for the dark baryon is
\begin{equation}
\begin{aligned}
 \mathcal B_\alpha&=\epsilon_{abc}(U^{aT}C\gamma_5D^b)U^c_\alpha\,,\\
 \langle0|\mathcal B_\alpha|B(p,s)\rangle&=\beta_{\rm D}\,u_\alpha(p,s)\,,
\end{aligned}
 \label{eq:baryon-overlap}
\end{equation}
where $a,b,c$ are dark-colour indices, $\alpha$ a Dirac index,
$C$ the charge-conjugation matrix, $|B(p,s)\rangle$ the one-baryon
state of momentum $p$ and spin $s$, and $u(p,s)$ its Dirac spinor.
The coefficient $\beta_{\rm D}$ measures the overlap of the three-quark
operator $\mathcal B$ with the physical baryon state.  We write $\beta_{\rm D}=\kappa_B\Lambda_{\rm D}^3$,
where $\kappa_B$ is a dimensionless non-perturbative coefficient,
chosen positive by convention.

Six dark quarks carry $N_{\rm D}=2$ and $Q_B=15$. We retain the
dimension-ten interaction
\begin{equation}
 \mathcal L_{\Delta N_{\rm D}=2}
 =-\frac{c_\Delta}{2\Lambda_*^6}
 S\mathcal B^TC\mathcal B+\mathrm{h.c.}\,,
 \label{eq:splitting-operator}
\end{equation}
where $c_\Delta$ is a dimensionless Wilson coefficient and $\Lambda_*$
the heavy matching scale. It violates the accidental $N_{\rm D}$ by
two units while respecting the gauge symmetry. Dimension ten is the
lowest dimension for this six-dark-quark splitting mechanism with the
stated fields, but different Lorentz, colour and flavour contractions
are possible.

Below the confinement scale, we match
\begin{equation}
 \mathcal B^TC\mathcal B
 \ \longrightarrow\ \kappa_\Delta\beta_{\rm D}^{2}\,\overline{B^c}B,
 \qquad B^c=C\bar B^T\,,
 \label{eq:hadronic-matching}
\end{equation}
where $\kappa_\Delta$ parametrises the six-quark matrix element relative
to $\beta_{\rm D}^2$, evaluated at the same renormalisation scale. We take it to
be of order unity.
After renormalisation-group evolution to the confinement scale and
matching onto the dark baryon, the operator in
Eq.~\eqref{eq:splitting-operator} generates a Majorana mass
\begin{equation}
\epsilon
=
\frac{c_\Delta U_\Delta\kappa_\Delta}{\sqrt{2}}\,
\frac{v_S\beta_{\rm D}^2}{\Lambda_*^6}\,,
\label{eq:majorana-mass}
\end{equation}
where $U_\Delta$ accounts for running between $\Lambda_*$ and
the confinement scale. We retain the operator in
Eq.~\eqref{eq:splitting-operator}, which gives equal left- and
right-handed Majorana masses, and rephase $B$ so that $\epsilon$ is
real and positive. Together with the Dirac mass $m_B$, it gives the
Majorana eigenstates
\begin{equation}
\chi_1=-\frac{\ii}{\sqrt{2}}(B-B^c),\qquad
\chi_2=\frac{1}{\sqrt{2}}(B+B^c)\,,
\label{eq:majorana-states}
\end{equation}
with masses $m_{1,2}=m_B\mp\epsilon$, so that $m_B=(m_1+m_2)/2$ and
$m_1\simeq m_B$. Their mass splitting is
\begin{equation}
\delta=m_2-m_1
=
\frac{\sqrt{2}\,
|c_\Delta U_\Delta\kappa_\Delta|\,v_S\beta_{\rm D}^2}
{\Lambda_*^6}\,.
\label{eq:splitting}
\end{equation}
Using $\beta_{\rm D}=\kappa_B\Lambda_{\rm D}^3$, the matching scale
required for a given splitting is
\begin{equation}
\begin{aligned}
\Lambda_*\simeq{}
13\,\mathrm{TeV}\,
\bigl(|c_\Delta U_\Delta\kappa_\Delta|\kappa_B^2\bigr)^{1/6}
&\left(\frac{v_S}{1\,\mathrm{TeV}}\right)^{1/6}\\
&
\frac{\Lambda_{\rm D}}{1\,\mathrm{TeV}}
\left(\frac{300\,\mathrm{keV}}{\delta}\right)^{1/6}\,.
\end{aligned}
\label{eq:matching-scale}
\end{equation}
For the benchmark splitting $\delta=300\,\mathrm{keV}$ we use below, with $\Lambda_{\rm D}\sim1\,\mathrm{TeV}$,
$v_S\sim0.6\,\mathrm{TeV}$ and order-one coefficients, this gives
$\Lambda_*\simeq12\,\mathrm{TeV}$.

\subsection{Vector portal and benchmark}\label{sec:portal}
In the Majorana basis, the dark baryon current is off diagonal:
\begin{equation}
\bar B\gamma^\mu B=\ii\bar\chi_2\gamma^\mu\chi_1\,.
\label{eq:transition-current}
\end{equation}
For momentum transfer $|q|\ll M_V$, vector exchange gives
\begin{equation}
 \mathcal L_{\rm eff}=-\ii G_N(\bar\chi_2\gamma_\mu\chi_1)
          (\bar p\gamma^\mu p+\bar n\gamma^\mu n)\,,
 \label{eq:contact}
\end{equation}
where $p,n$ are proton and neutron fields and
\begin{equation}
 G_N=\frac{Q_B(B) g_B^2}{M_V^2},\qquad
 \sigma_p^0=\frac{\mu_p^2G_N^2}{\pi}\,,
 \label{eq:portal}
\end{equation}
where
$\mu_p=m_1m_p/(m_1+m_p)$, and both nucleons carry unit baryon charge.
The neutron-to-proton coupling ratio is therefore $f_n/f_p=1$ up to the kinetic mixing we discuss below.
The quantity $\sigma_p^0$ is the elastic, zero-momentum reference
normalisation.
Kinetic mixing,
$\mathcal L_{\rm mix}=-\tfrac{\epsilon_{YB}}2F^Y_{\mu\nu}V^{\mu\nu}$,
induces lepton couplings and shifts the proton and neutron couplings.
For our charges, $\operatorname{Tr}(YQ_B)=0$, so the one-loop additive
running cancels above all spectator thresholds. Threshold corrections
and running below those thresholds can generate non-zero mixing.
We assume it is small enough to neglect in the benchmark and therefore
$f_n/f_p=1$.

We use the following benchmark
\begin{equation}
 \begin{split}
m_1&=1.0\,\mathrm{TeV},\qquad\delta=300\,\mathrm{keV}\,,\\
 M_V&=1.44\,\mathrm{TeV},\qquad g_B=0.146\,,\\
 v_\Phi&=1.0\,\mathrm{TeV},\qquad v_S\simeq0.63\,\mathrm{TeV}\,.
 \end{split}
 \label{eq:benchmark}
\end{equation}
This gives $M_V/g_B=9.86\,\mathrm{TeV}$,
$g_q=0.049$ and $\sigma_p^0=6.5\times10^{-43}\,\mathrm{cm^2}$.
The benchmark lies inside LZ's approximate two-sided 90\% interval
$(3.6\times10^{-43},7.2\times10^{-42})\,\mathrm{cm^2}$, digitised from
Supplemental Fig.~S7 of Ref.~\cite{LZ:2026axp}.
It is an illustrative point within that interval, not a best-fit
determination of the portal strength: since
$\sigma_p^0\propto(g_B/M_V)^4$, the interval constrains $M_V/g_B$ only to
within a factor of about two. 
At $|q|\ll M_V$ and leading order in the dark matter velocity $v\sim10^{-3}$,
Eq.~\eqref{eq:contact} is the isoscalar spin-independent operator
$O_1^s$ of the non-relativistic effective theory in which LZ reports its
results and which \textsc{WimPyDD} takes as input. We calculate the
endothermic benchmark assuming that all of the local dark matter is in
$\chi_1$, with $f_1\equiv\rho_{\chi_1}/\rho_{\rm DM}\simeq1$,
and use LZ's isoscalar cross-section convention.

\section{The xenon recoil spectrum}
\label{sec:spectrum}
For dark matter of mass $m_1$ scattering off a nucleus of mass
$m_N$, with reduced mass $\mu_{\chi N}=m_1m_N/(m_1+m_N)$, the minimum
speed producing a recoil energy $E_R$ is
\begin{align}
 v_{\min}(E_R)&=\frac{m_NE_R/\mu_{\chi N}+\delta}{\sqrt{2m_NE_R}}\,,\nonumber\\
 E_R^\star&=\frac{\mu_{\chi N}}{m_N}\delta\,,\qquad
 v_{\rm thr}=v_{\min}(E_R^\star)=\sqrt{\frac{2\delta}{\mu_{\chi N}}}\,,
 \label{eq:kinematics}
\end{align}
where $E_R^\star$ is the recoil energy at which $v_{\min}$ is
smallest and $v_{\rm thr}$ the threshold speed.
Two properties of the model matter for LZ.
The first is kinematic. In the adopted truncated Maxwellian halo,
Galactic-frame speeds are restricted to $v\le v_{\rm esc}$, with the Galactic escape
speed $v_{\rm esc}\simeq544\,\mathrm{km\,s^{-1}}$. The Earth moves through
the halo at $v_E\simeq250\,\mathrm{km\,s^{-1}}$, so the fastest particle
that can reach the detector arrives at
$v_{\max}=v_{\rm esc}+v_E\simeq794\,\mathrm{km\,s^{-1}}$. A recoil energy
$E_R$ is therefore possible only if $v_{\min}(E_R)\le v_{\max}$. At the
benchmark this allows xenon recoils only in the range
$98\lesssim E_R/\mathrm{keV}\lesssim727$ for $m_N=122\,\mathrm{GeV}$. The second property is
the rate. It is set by $\sigma_p^0$, which through Eq.~\eqref{eq:portal}
depends only on $M_V/g_B$.
For $m_N=122\,\mathrm{GeV}$, Eq.~\eqref{eq:benchmark} gives
$E_R^\star=267\,\mathrm{keV}$ and $v_{\rm thr}=704\,\mathrm{km\,s^{-1}}$,
so the benchmark is kinematically accessible but probes the high-speed
tail of the halo, and the observed energy lies close to the minimum of
$v_{\min}(E_R)$.

For natural xenon, the differential rate per unit detector mass, at
leading order in $q^2/M_V^2$, is
\begin{equation}
 \frac{dR}{dE_R}=\frac{\rho_1\sigma_p^0}{2m_1\mu_p^2}
 \sum_i\xi_i A_i^2F_i^2(q_i)\,\eta(v_{\min,i})\,,
 \label{eq:rate}
\end{equation}
where $\xi_i$ are isotope mass fractions, $A_i$ and $m_{N_i}$ are the
isotope mass numbers and nuclear masses, and
$q_i=|\mathbf q_i|=\sqrt{2m_{N_i}E_R}$ is the momentum transfer;
$F_i^2$ denotes the isoscalar nuclear mass response normalised to unity
at zero momentum transfer, and
$\eta(v)=\int_{|\mathbf u|>v}d^3u\,f_{\rm lab}(\mathbf u)/|\mathbf u|$ is
the mean inverse speed of halo particles faster than $v$, evaluated at
$v=v_{\min,i}(E_R)$. Here $f_{\rm lab}$ is the laboratory-frame velocity
distribution, normalised to $\int d^3u\,f_{\rm lab}(\mathbf u)=1$.
The incoming-state density is $\rho_1=f_1\rho_{\rm DM}$ with $f_1=1$.
We use a truncated Maxwellian with $v_0=238\,\mathrm{km\,s^{-1}}$,
$v_{\rm esc}=544\,\mathrm{km\,s^{-1}}$ and
$\rho_{\rm DM}=0.3\,\mathrm{GeV\,cm^{-3}}$ \cite{Baxter:2021pqo}, with
the Earth speed fixed at $250.2\,\mathrm{km\,s^{-1}}$, close to its
annual mean. Because the
benchmark sits close to the kinematic threshold, the rate depends
sensitively on the assumed high-speed halo distribution.

\begin{figure*}[t]
 \centering
 \includegraphics[width=0.96\textwidth]{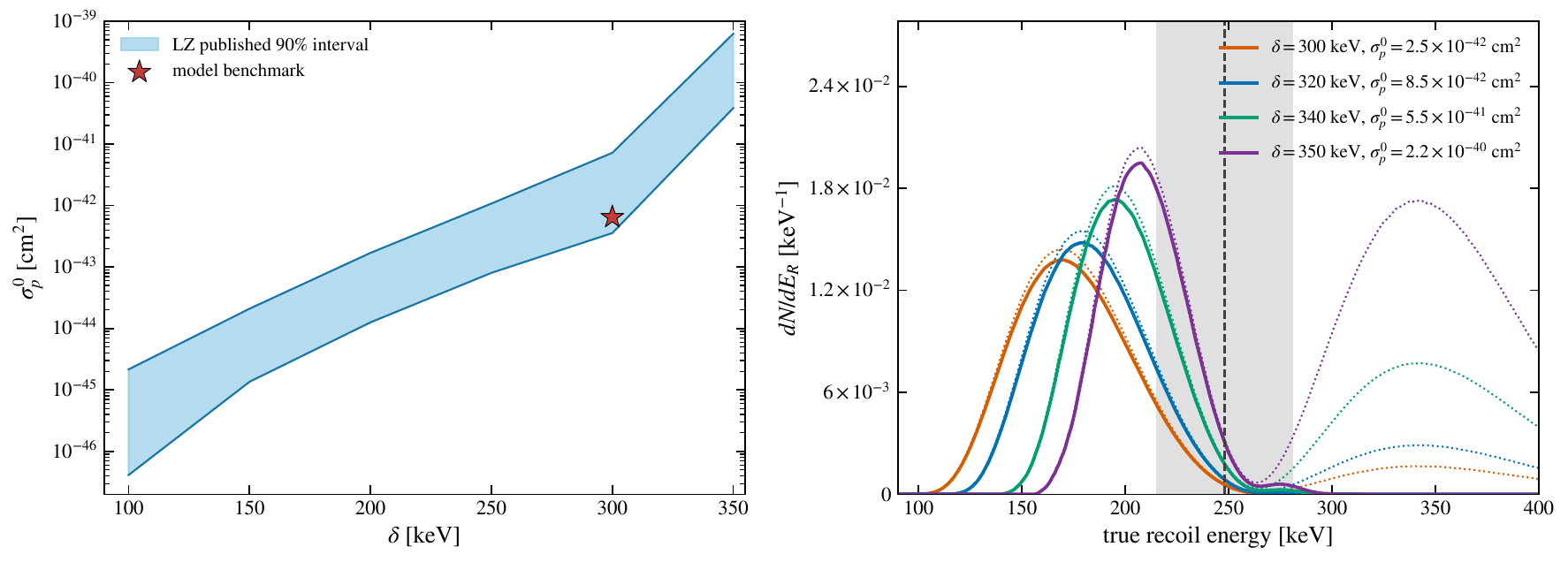}
 \caption{Left: the model benchmark (star) against LZ's published two-sided
 90\% interval on the isoscalar spin-independent cross section for a
 $1\,\mathrm{TeV}$ particle scattering through $O_1^s$, as a function of
 the splitting $\delta$. Right: recoil spectra in true recoil energy for
 $2.84$ tonne-years, calculated with \textsc{WimPyDD}'s supplied xenon
 responses, for four splittings. Each solid curve is the spectrum after
 folding with the digitised LZ nuclear-recoil efficiency, at the cross
 section given in the legend, chosen so that it integrates to one
 accepted event; the dotted curves are the same spectra before the
 efficiency. At fixed $M_V=1.44\,\mathrm{TeV}$ and $v_\Phi=1\,\mathrm{TeV}$
 the points with $\delta\ge320\,\mathrm{keV}$ exceed the perturbativity
 criterion of Sec.~\ref{sec:spectrum}, and the $350\,\mathrm{keV}$ point is
 incompatible with the vector-mass relation. The dashed line marks the reconstructed energy of the LZ
 event and the grey band its $\pm33\,\mathrm{keV}$ combined uncertainty,
 shown for orientation only, since no energy resolution is applied to the
 curves.  }
 \label{fig:spectra}
\end{figure*}

We calculate the recoil spectrum with \textsc{WimPyDD}~1.6.2
\cite{Jeong:2021bbi}, using its supplied xenon nuclear responses and
LZ's nuclear-recoil efficiency, digitised as a function of true recoil
energy from Ref.~\cite{LZ:2026axp}. We apply no additional energy
smearing and assume no visible energy from $\chi_2$ de-excitation
inside the detector. These inputs determine the spectra and event
yields below; our statistical comparison uses LZ's published interval,
which incorporates their nuclear and detector responses.
Before acceptance, the benchmark spectrum peaks near
$169\,\mathrm{keV}$ and has a diffraction minimum near
$266\,\mathrm{keV}$. The predicted rate near the observed
$248\,\mathrm{keV}$ recoil is therefore sensitive to the nuclear
response. With the adopted inputs, the benchmark yields approximately
$0.32$ events before acceptance and $0.26$ after acceptance for
$2.84$ tonne-years.

The left panel of Fig.~\ref{fig:spectra} shows LZ's published
two-sided 90\% interval on the isoscalar cross section as a function
of the splitting; our benchmark lies within this interval at
$\delta=300\,\mathrm{keV}$. The right panel shows
the recoil spectra before and after acceptance for four splittings,
with each cross section adjusted to yield one accepted event.
The $\delta=300\,\mathrm{keV}$ curve therefore has the benchmark
shape but a different normalisation. The dashed vertical line and
grey band indicate the reconstructed LZ event energy and its
$\pm33\,\mathrm{keV}$ uncertainty. These provide a visual comparison
only, since the spectra are not convolved with an energy-resolution
model. Larger splittings shift the spectral peak towards the observed
energy, but require rapidly increasing cross sections.

For this scan we fix $M_V=1.44\,\mathrm{TeV}$ and
$v_\Phi=1\,\mathrm{TeV}$. A larger splitting leaves fewer halo
particles able to scatter, so the required $g_B$ grows quickly.
Taking $(15g_B)^2/4\pi\lesssim1$ as the perturbativity criterion,
with $|Q_B(S)|=15$ the largest charge in the model, this quantity
is $0.4$ at the benchmark, $0.7$ at the one-event point for
$\delta=300\,\mathrm{keV}$ and $1.4$ at $320\,\mathrm{keV}$.
At $340\,\mathrm{keV}$ the dark-baryon coupling also becomes
non-perturbative, and at $350\,\mathrm{keV}$ the required
$M_V/g_B\simeq2.3\,\mathrm{TeV}$ is below the minimum $3v_\Phi$
set by Eq.~\eqref{eq:vector-mass}. These considerations motivate
our choice of $\delta=300\,\mathrm{keV}$.

\section{Constraints and conclusions}
\label{sec:constraints}
\textit{High-energy neutrinos from the Sun.}---Dark matter
captured by the Sun can annihilate within it. The annihilation
products can generate high-energy neutrinos through decays and
hadronic cascades. Searches for this flux constrain the capture
rate and hence the scattering cross section, subject to assumptions
about the annihilation rate and final states
\cite{Pospelov:2026sol,Nguyen:2026lui}. Our interaction
matches the isoscalar $O_1^s$ case of Ref.~\cite{Nguyen:2026lui}. At
$m_1=1\,\mathrm{TeV}$ and $\delta=300\,\mathrm{keV}$, and for $f_1=1$,
that study quotes a 90\% confidence limit
$\sigma_p^0<9.9\times10^{-43}\,\mathrm{cm^2}$ assuming annihilation
entirely to $b\bar b$, with stronger limits for $W^+W^-$, $\tau^+\tau^-$
and $\nu\bar\nu$ final states. These limits assume the specified
thermalised solar distribution and annihilation rate and the latter
three channels do not arise at tree level for our leptophobic
SM-singlet baryon. In our model the captured baryons annihilate into dark hadrons,
which decay to quarks. The neutrino yield of this cascade depends
on the dark-hadron spectrum and lifetimes, so the $b\bar b$ limit
serves only as a reference value. If the dark hadrons decay predominantly
to light quarks within the dense solar medium, energy loss of
secondary charged mesons can suppress the high-energy neutrino
yield, which can weaken the bound.

\textit{Accelerators.}---The vector $V$ can be produced
through $q\bar q\to V$ and appear as a dijet resonance. The ATLAS
search \cite{ATLAS:2025okg} constrains a leptophobic axial-vector
mediator with universal quark couplings and negligible dark decays.
Neglecting quark masses and interference, vector and axial couplings
give the same leading-order production rate and angular distribution.
For a narrow resonance with comparable acceptance and line shape,
the bound therefore applies approximately as
$g_q\sqrt{B_{\rm SM}}\lesssim g_q^{95}(M_V)$, where $B_{\rm SM}$
is the branching fraction into SM quarks and $g_q^{95}(M_V)$
is the ATLAS 95\% confidence-level upper limit. At
$M_V=1.44\,\mathrm{TeV}$, the translation of Fig.~9(b) of
Ref.~\cite{ATLAS:2025okg} used in Ref.~\cite{Du:2026lpa} gives
$g_q^{95}\simeq0.084$, above our benchmark $g_q\simeq0.05$
even for $B_{\rm SM}=1$.

Dark-hadron decays can reduce the direct dijet signal and produce
missing energy or displaced or semi-visible jets, depending on
the spectrum and lifetimes. A background-limited extrapolation,
$g_q^{95}\propto\mathcal{L}^{-1/4}$, where $\mathcal{L}$ is the
integrated luminosity, suggests a dijet reach of
$g_q^{95}\sim0.04$ at $\mathcal{L}=3\,\mathrm{ab}^{-1}$,
which could probe the benchmark at the HL-LHC if decays
to SM quarks dominate. Identifying a resonance with $V$ would
determine its mass; the LZ cross-section interval would then
constrain its quark coupling. Comparing the predicted and measured
dijet rates, accounting for the branching fraction, would test
whether the same interaction explains the LZ recoil.

\textit{Abundance and excited states.}---We assume that the
local dark matter density is predominantly $\chi_1$. With the same 
acceptance, an excited fraction $f_2\equiv\rho_{\chi_2}/\rho_{\rm DM}$
gives an exothermic recoil rate about $10^3f_2$ times the endothermic
one. Thus $f_2\sim10^{-3}$ would already contribute comparably, and the
interpretation requires $f_2<10^{-3}$. Dark-number-conserving
production of $B$ and $\bar{B}$ gives equal $\chi_1$ and $\chi_2$
populations. 
The excited state can instead decay. For $m_a<\delta$, the phase of $S$
in the Majorana mass allows $\chi_2\to\chi_1a$, with a lifetime
$\tau\simeq2.6\times10^{-6}\,\mathrm{s}$ for $m_a\ll\delta$.
This decay can deplete $\chi_2$ before
nucleosynthesis once inverse decays become ineffective. It requires a
strongly suppressed phase-locking interaction, and the light
pseudoscalar must satisfy astrophysical constraints. We assume that
$a$ escapes the detector without a detectable energy deposit.
For the total abundance, the geometric estimate
$\sigma v\sim\pi v/\Lambda_{\rm D}^2\sim10^{-23}\,\mathrm{cm^3\,s^{-1}}$
at $v\sim0.3$ and $\Lambda_{\rm D}\sim1\,\mathrm{TeV}$ suggests an
underabundant symmetric thermal relic. A full abundance calculation must
include the thermal history and decays of the lighter dark hadrons
\cite{Dondi:2019olm,Howard:2021ohe}. 
An asymmetric or non-thermal
origin must also account for the oscillations induced by the Majorana
splitting \cite{Ibe:2019yra,Slatyer:2015jla}.

\textit{Summary.}---We consider a composite dark baryon with a dimension-ten Majorana
splitting that realises endothermic scattering through a vector portal. Our benchmark, a $1\,\mathrm{TeV}$ dark baryon with a $300\,\mathrm{keV}$
splitting between its two Majorana states, has
$\sigma_p^0\simeq6.5\times10^{-43}\,\mathrm{cm^2}$, lies within LZ's
published isoscalar two-sided 90\% interval and gives about $0.26$
accepted events. The splitting suppresses low-energy recoils and
the portal fixes the normalisation. We have examined the constraints
most likely to challenge this picture: solar-neutrino searches, dijet
resonance limits and the survival of the excited state. The benchmark
sits below the published solar bound even in its most conservative
form, below the dijet limit even if $V$ decays only to quarks, and the
model itself provides a fast decay of $\chi_2$ through a light
pseudoscalar. The interpretation remains possible, but establishing its
viability requires dedicated calculations.
Further recoil data can test the interpretation directly, and the
HL-LHC can probe the mediator through dijet resonances or, if dark
decays dominate, through missing-energy and displaced-jet signatures.

\begin{acknowledgments}
The work of F.S. is partially supported by the Carlsberg Foundation,
Semper Ardens grant CF22-0922. F.S. and J.T. acknowledge support from the
Royal Society under grant IES\textbackslash{}R1\textbackslash{}261011.
During the preparation of this work, we used ChatGPT and Claude to assist with literature searches, the development of Python calculations using WimPyDD, and manuscript revision. The authors take full responsibility for the analysis and final manuscript.
\end{acknowledgments}
\bibliography{references}

\end{document}